\documentstyle[prd,aps,epsfig,floats]{revtex}
\begin{document}  \draft
\wideabs{
\title{Warming or cooling from a random walk process in the temperature}

\author{Bernd A. Berg}

\address{Department of Physics, Florida State University, 
         Tallahassee, FL 32306-4350} 

\date{\today } 

\maketitle
\begin{abstract}
A simple 3-parameter random walk model for monthly fluctuations 
$\triangle T$ of a temperature $T$ is introduced. Applied to a 
time range of 170 years, temperature fluctuations of the model produce 
for about 14\% of the runs warming that exceeds the observed 
global warming of the earth surface temperature from 1850 to 2019. 
On the other hand, there is a 50\% likelihood for runs of our model 
resulting in cooling. If a similar random walk process can be used 
as an effective model for fluctuations of the global earth surface 
temperature, effects due to internal and external forcing could 
be considerably over- or underestimated.
\end{abstract}
} \narrowtext


\begin{figure}[th] \begin{center} 
\epsfig{figure=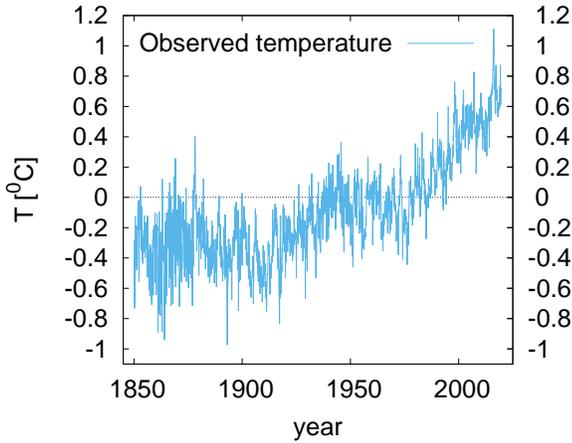,
width=\columnwidth} \caption{Global monthly temperature averages from 
1850 to 2019 (Hadley Center). \vspace{-5mm} \label{fig_HadCRUT4}} 
\end{center} \end{figure} 

Global warming has become a subject of major research efforts 
\cite{IPCC}. Figure~\ref{fig_HadCRUT4} relies on a data set of 
the UK Met Office Hadley Center \cite{HadCRUT4} and depicts monthly 
estimates for the global earth surface temperature from 1850/01 to 
2019/09.  From about 1975 on a global warming trend is clearly visible 
and consistent with satellite based monthly temperature estimates, which 
are available from 1979 on \cite{Roy}. The zero reference line adapted
in Fig.~\ref{fig_HadCRUT4} is close to the mean value of all shown data, 
$\overline{T}=-0.088\,[^0\rm C]$. In this paper all temperature 
references (in Celsius) are with respect to the zero line and not with 
respect to the Celsius scale.

The global temperature is chosen for our considerations, because one
expects significantly less seasonal variations than, for instance, for
the temperatures of the northern or southern hemisphere. Though some 
asymmetry due to the seasons remains, it is not immediately visible 
from the graph of Fig.~\ref{fig_HadCRUT4}. Instead, one realizes 
already at a first glance that the temperature curve is not smooth, 
but fluctuates heavily from month to month. How does this happen?
The temperatures of Fig.~\ref{fig_HadCRUT4} are weighted averages 
over measurements in a narrow band close to the surface of the earth. 
Energy exchanges in the horizontal directions are balanced to zero
by energy conservation, while there can be mismatches of incoming 
and outgoing energies in the vertical directions.

Heating comes mainly from the radiation of the sun. About 30\% of 
the sun's radiation gets immediately reflected back into outer space.
What is left over heats the ground, the oceans and the atmosphere. 
Ultimately the heat escapes in the form of mostly infrared radiation. To 
avoid continuous heating or cooling, the energy of the incoming radiation 
has to agree in average with that of the outgoing radiation. Due to 
statistical fluctuations this balance does not hold at every instance.
For example, if there are clouds at daytime, more sunlight will 
immediately be reflected back into space than on a clear day. The 
effect exceeds the trapping of infrared radiation by the clouds and 
it will be cooler than on a sunny day under otherwise similar conditions. 
At night the opposite is true, because only the trapping effect remains.
Heat exchange with the oceans and the earth surface has also random 
components, and so on. 

It is shown here that 
accidental fluctuations of a random walk process with mean $\left< 
\triangle T\right> =0$, i.e., a process that has no preference for 
increasing or decreasing the temperature, can exhibit similar 
temperature drifts as shown in Fig.~\ref{fig_HadCRUT4}.

\begin{figure}[th] \begin{center} 
\epsfig{figure=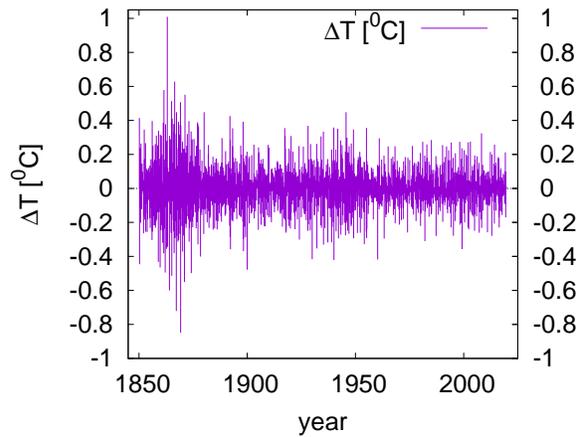,
width=\columnwidth} \caption{Global monthly temperature fluctuations.
\vspace{-8mm} \label{fig_monthly}} \end{center} \end{figure} 

Let us label the monthly temperatures of Fig.~\ref{fig_HadCRUT4} by
$T(i),\ i=1,\dots,2037$. The corresponding dates are approximately 
given by the equation
\begin{eqnarray} \label{eq_date}
  {\rm Date} = (1850 + i/12)\,{\rm years}\,,~~~i=1,\dots, 2037\,.
\end{eqnarray}
Monthly temperature fluctuations are then defined by
\begin{eqnarray} \label{eq_dT}
  \triangle T(i) = T(i) - T(i-1)\,,~~~i=2,\dots, 2037\,,
\end{eqnarray}
and their time series is depicted in Fig.~\ref{fig_monthly}.  On closer 
inspection of the data one finds that the 13 largest monthly temperature 
fluctuations all fall into the time period before 1900 although far 
more data points exist from 1900 on. This may be a real effect or due 
to the larger uncertainties of the older data. In either case the older
data enlarge substantially the standard deviation of the empirical 
distribution
\begin{eqnarray} \label{eq_sigma}
  \sigma_e = \sqrt{\frac{1}{2035}\,\sum_{i=2}^{2037}\left(\triangle 
  T(i) - \overline{\triangle T}\right)^2} = 0.138\ [^0\rm C]\,.
\end{eqnarray}
Here $\overline{\triangle T}=0.000695\,[^0\rm C]$ is the mean value of 
the monthly temperature fluctuations:
\begin{eqnarray} \label{eq_T}
  \overline{\triangle T} = \sum_{i=2}^{2037} \frac{\triangle T(i)}{
  2036} = \frac{T(2037)-T(1)}{2036} = \frac{1.415\ [^0\rm C]}{2036}
\end{eqnarray}
where $T(1)=-0.7\,\rm [^0C]$ and $T(2037)=0.715\,\rm [^0\rm C]$ are 
the temperatures for the first and last month considered.

\begin{figure}[th] \begin{center} 
\epsfig{figure=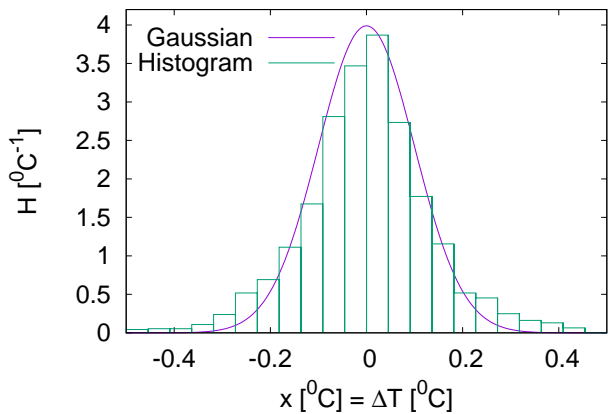, width=\columnwidth} \caption{Probability 
density of global monthly temperature fluctuations. \label{fig_probden}} 
\vspace{-5mm} \end{center} \end{figure} 

The probability density of $\triangle T$ is depicted in 
Fig.~\ref{fig_probden} in form of a histogram together with 
the Gaussian probability density 
\begin{eqnarray} \label{eq_gauss}
 H = \frac{1}{\sigma_g\sqrt{2\pi}}\,\exp\left[-\frac{1}{2}\,\left(
  \frac{x}{\sigma_g}\right)^2\right]\,.
\end{eqnarray}
of standard deviation $\sigma_g =0.1\ [^0\rm C]$. This Gaussian fits 
the peak of the histogram quite well, while it ignores the outliers. 
For our purposes it is sufficient that the Gaussian creates temperature 
fluctuations in the more recently observed range of natural variability.
Note that the mean value of the Gaussian is chosen to be $\widehat{x}
=\left<x\right>=0$, whereas $\overline{\triangle T}$ is non-zero 
(\ref{eq_T}) for the empirical $\triangle T$ distribution. Further, 
Gaussian fluctuations are statistically independent, whereas this
is not expected for the observed fluctuations. The Gaussian form 
is chosen for simplicity. Similar arguments could be made using
the empirical histogram directly to generate uncorrelated or
correlated random walk updates in some kind of bootstrap approach.

Although the standard deviation $\sigma_g=0.1\, [^0\rm C]$ is chosen 
smaller than $\sigma_e$ (\ref{eq_sigma}) of the observed fluctuations, 
Gaussian temperature fluctuations with $\sigma_g=0.1\,[^0\rm C]$ are 
too large to comply with the observed temperature drift over the last 
170 years. This is shown next by investigating a sample of $n_{rpt} =
1001$ Gaussian random walks, each of $2040$ steps generated with the 
probability density~(\ref{eq_gauss}). For all simulations reported 
here we use Marsaglia random numbers and other software from 
\cite{Berg}.

\begin{figure}[th] \begin{center}\epsfig{figure=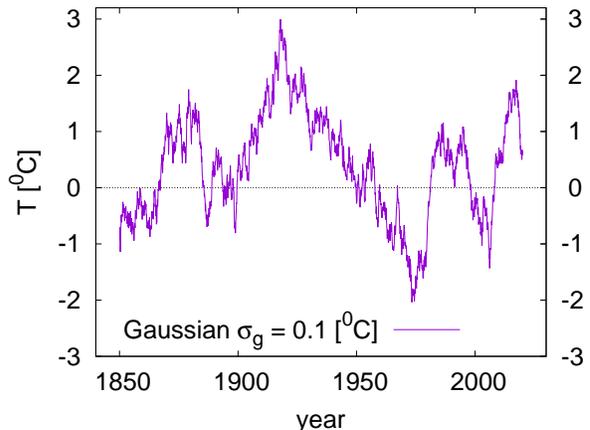,
width=\columnwidth} \caption{Typical temperature random walk for
Gaussian fluctuations with $\sigma_g=0.1\,\rm [ ^0C]$.  
\label{fig_gau10}} \vspace{-5mm} \end{center} \end{figure} 

For each of the random walks the minimum $T_{\min}(i_{rpt})$ and the 
maximum $T_{\max}(i_{rpt})$, $i_{rpt}=1, \dots,1001$, temperature were 
recorded. The differences $T_{\max}(i_{rpt})-T_{\min}(i_{rpt})$ are then 
found in the range $2.7\,[ ^0\rm C]$ to $18.3\,[ ^0\rm C]$ with a median 
of $6.7\,[ ^0C]$. An example with $T_{\max}-T_{\min}=5.0\, ^0C$ is given 
in Fig.~\ref{fig_gau10}. Over the considered time period temperature 
excursions were far too large to resemble those of 
Fig.~\ref{fig_HadCRUT4}. They increase proportional to $\sigma_g
\sqrt{n}$ ($n=2040$), and $\sigma_g=0.1\,[^0C]$ is too large to allow 
the random walk to stay within the range given by the observations of 
Fig.~\ref{fig_HadCRUT4}. On the other hand, we cannot change $\sigma_g$ 
much because of the $\triangle T$ time series of Fig.~\ref{fig_monthly}. 
The requirement $\widehat{x}=0$ is not sufficient to prevent a run-away 
to very hot or cool temperatures. Some kind of ``thermostat'' is needed, 
which drives the random walk back to the neighborhood of $\widehat{x}=0$ 
without changing its variance.

\begin{figure}[tb] \begin{center} \epsfig{figure=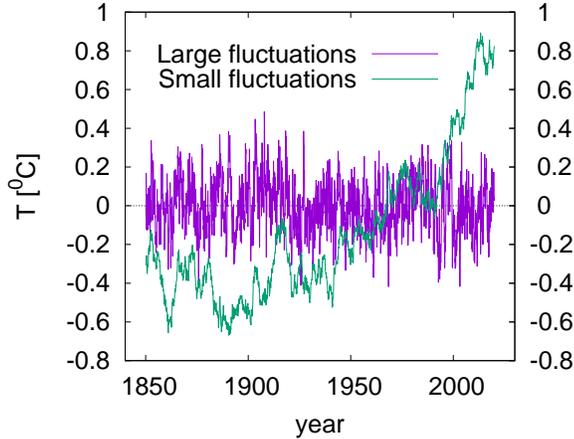,
width=\columnwidth} \caption{Examples of large and small fluctuations 
of monthly temperature changes from random walks. \vspace{-5mm}
\label{fig_modgau}} \end{center}\end{figure} 

We achieve this by turning a proposed update $\triangle T=x$ into 
$\triangle T=-x$ with a suitable likelihood. Due to $\widehat{x}=0$ the 
variance $\sigma^2_g=\left<x^2\right>$ is invariant under $x\to -x$. For 
the construction of a suitable likelihood we introduce the probabilities
\begin{eqnarray}
  p=\int_{-\infty}^{a\,T(i-1)}\ g(x)\,dx~~{\rm and}~~
  q=\int_{a\,T(i-1)}^{\infty}\ g(x)\,dx\,,
\end{eqnarray}
$p+q=1$, where $g(x)$ is the Gaussian probability density 
(\ref{eq_gauss}) with $\sigma_g=1$ and $a>0$ a free parameter. 
The update will now be $\triangle T(i) = \pm \left|x\right|$ where 
the sign is determined as follows: 
\begin{eqnarray} \label{deltaT}
  \triangle T(i) = \cases{-|x|\ {\rm  with\ probability}\ p,\cr
                          +|x|\ {\rm  with\ probability}\ q. } 
\end{eqnarray}
In each case the larger of the probabilities $p$ and $q$ drives the 
random walk defined by
\begin{eqnarray}
  T(i-1)\to T(i) = T(i-1) + \triangle T(i) 
\end{eqnarray}
closer to zero without changing the variance. These fluctuations are
no longer statistically independent. In the following they are called
``modified'' Gaussian fluctuations. The large fluctuations around zero 
of Fig.~\ref{fig_modgau} exhibit a typical example of a random walk 
for the temperatures $T(i)$ obtained from our modified Gaussian 
fluctuations. It starts with $T(1)=0$, and $a=4$ is used for the 
free parameter.

To reproduce the temperature increase seen in Fig.~\ref{fig_HadCRUT4} 
one may now add a smooth curve representing systematic causes like 
internal and external forcings. However, they are not the subject of 
this paper. Here we investigate whether a purely statistical random 
walk model can create similar temperature drifts. The idea is to
achieve this by adding to the large modified Gaussian fluctuations
ordinary Gaussian fluctuations with a standard deviation $\sigma'_g
\ll 0.1\,[^0C]$ scaled so that a temperature difference like the one 
depicted in Fig.~\ref{fig_HadCRUT4} is within reach over the given 
time range. A good choice for this are Gaussian random walks with 
$\sigma'_g=0.02\,[^0C]$. An example is given by the ``small 
fluctuations'' of Fig.~\ref{fig_modgau}. This assumes that in 
the range of presently relevant temperatures there is no effective 
backdriving mechanism for these small temperature fluctuations.

\begin{figure}[th]\begin{center}\epsfig{figure=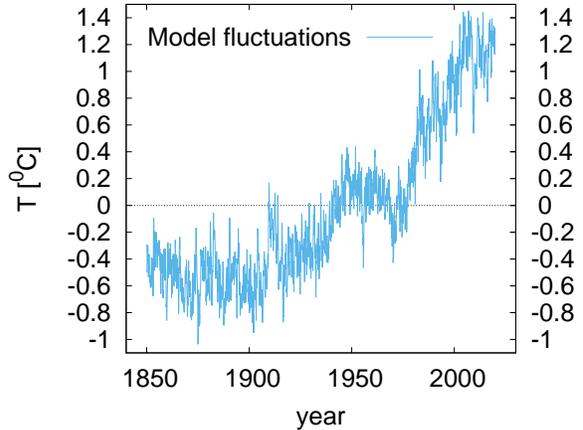,
width=\columnwidth} \caption{Special case random model monthly 
temperature changes. Compare with Fig.~1. \label{fig_tstemp}} 
\end{center}\end{figure} 

Simply adding a modified Gaussian walk with $\sigma_g=0.1\,[^0C]$ 
and a Gaussian random walk with $\sigma'_g=0.02\,[^0C]$ defines our 
model for temperature fluctuations. The example obtained by adding 
the large and small fluctuations of Fig.~\ref{fig_modgau} is shown 
in Fig.~\ref{fig_tstemp}. The similarity with the observed temperature 
fluctuations of Fig.~\ref{fig_HadCRUT4} is almost unbelievable. 
Therefore, I have posted the program that creates the temperature 
fluctuations of Fig.~\ref{fig_tstemp} on my website 
www.hep.fsu.edu/$\sim$berg/research/research.html ,


\noindent so that readers can verify by themselves that 
Fig.~\ref{fig_tstemp} is indeed a result of the random process
of our model. Download the archive file temperature.tgz, expand 
it, and follow the instructions of the readme.txt file.

The temperature increase from 1975 to 2019 is in Fig.~\ref{fig_tstemp} 
slightly stronger than in Fig.~\ref{fig_HadCRUT4}. Based on a sample 
of 10001 random walks of length $n=2040$ generated with our model one 
finds for 13.7\% a temperature increase that exceeds the observed 
increase. In the following we analyze this sample further.

We define the initial and final temperatures, $T_1$ and $T_2$ 
respectively, of each random walk as averages over the 
initial and final 2.5\% of the data:
\begin{eqnarray} \label{T1T2}
  T_1 = \frac{1}{n_1}\sum_{i=1}^{n_1}   T(i)~~{\rm and}~~
  T_2 = \frac{1}{n_1}\sum_{i=n+1-n_1}^n T(i)\,,
\end{eqnarray}
where $n_1=0.025\,n=51$. The averaging procedure (\ref{T1T2}), instead 
of just $T(1)$, $T(n)$, is used to reduce the dependence on accidental 
fluctuations. For the observed ($o$) monthly temperatures ($n=2037$ then)
of Ref.~\cite{HadCRUT4} one finds 
\begin{eqnarray} \label{T1T2observed}
  T_1^{\rm o} = -0.2780\,[ ^0\rm C ]\,,~~
 (T_2^o-T_1^o) = 0.9904\,[^0\rm C]\,.
\end{eqnarray}
For our model we sorted the $n_{rpt}=10001$ generated temperature
random walks in increasing order with respect to $(T_2-T_1)(i_{rpt})$. 
Of those $1\,372$ exceeded the value of $0.99\,[^0\rm C]$. 

\begin{figure}[th] \begin{center} \epsfig{figure=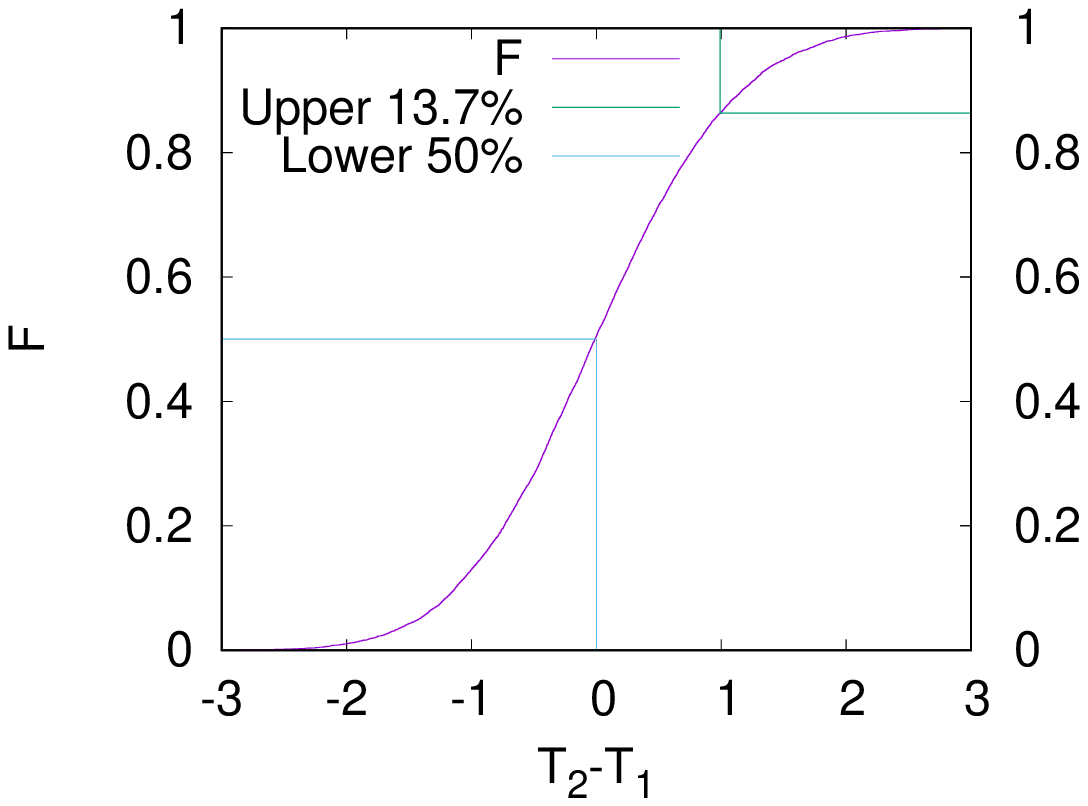,
width=\columnwidth} 
\caption{Cumulative distribution function $F(T_2-T_1)$ of the random 
walk model temperature increase $T_2-T_1$. \label{fig_dftemp}} 
\epsfig{figure=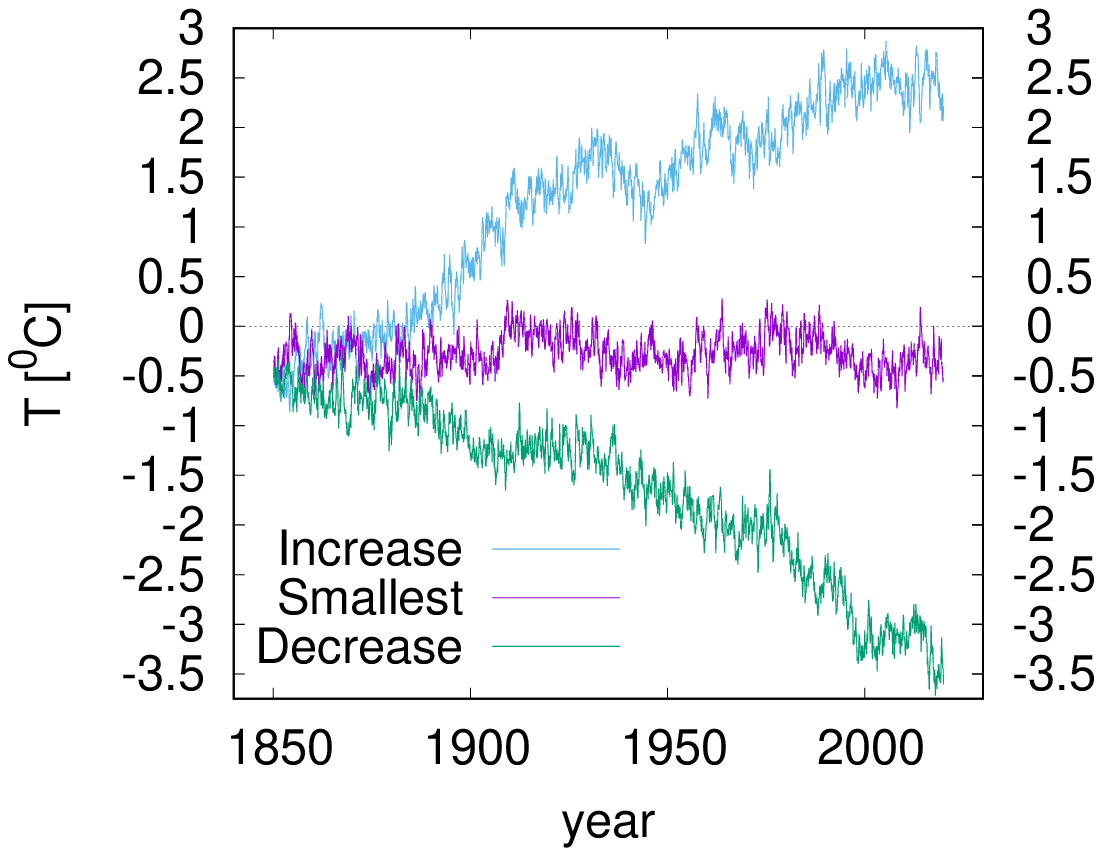,
width=\columnwidth} \caption{Examples of random walks from our model:
Largest temperature increase $T_2-T_1>0$, smallest absolute value for 
its maximum minus its minimum temperature, largest temperature decrease 
$T_2-T_1<0$.  \label{fig_tstemps}} \end{center}\end{figure} 

Figure~\ref{fig_dftemp} shows the cumulative distribution 
function $F(T_2-T_1)$ from our simulation. In the upper right the
13.7\% q-tile is indicated for which the warming is greater-equal
than the observed warming. In the lower left we have the 50\% range 
of values for which there is cooling by the random process.  
Figure~\ref{fig_tstemps} depicts three extreme cases from our 
sample of 10001 random walks.

In conclusion, the simple stochastic model of this paper exhibits the
variability needed to describe the global temperature changes for
the time period over which global temperature records exist. 

The large monthly temperature fluctuations of Fig.~\ref{fig_monthly} 
are an observed fact of the real world. In Fig.~\ref{fig_probden} the 
central part of their empirical probability density is approximated 
by a Gaussian with a variance suggested by the empirical data. However, 
Gaussian fluctuations with this variance lead to temperature excursions 
which are far too large to comply with observations. An example is shown 
in Fig.~\ref{fig_gau10}. This problem is overcome by correlating moves 
in a suitable statistical way (\ref{deltaT}), where the particular 
mechanism used is not really of importance for our present discussion.

The challenge remaining is to exclude the existence of subleading small 
fluctuations which are effectively uncorrelated over a temperature range 
larger or equal to that of Fig.~\ref{fig_HadCRUT4}. In our illustration 
subleading Gaussian fluctuations with a variance of 2.5\% of the 
variance (20\% of the standard deviation) of the leading fluctuations 
are chosen. That is about the maximum allowed. Substantially larger 
subleading Gaussian fluctuations are excluded by the observations. 

The assumed subleading fluctuations make it impossible to identify the 
causes for the temperature increase in Fig.~\ref{fig_HadCRUT4}. With 
a probability of almost 14\% one would have a natural temperature 
increase that is larger than the observed one, while with a probability 
of 50\% causal reasons like external forcing by an increase of the 
CO$_2$ contents of the atmosphere would have to be even larger than 
indicated by the observations. Phenomena like the medieval warm period 
and the little ice age would need no causal explanations anymore.

These problems would be demagnified for subleading Gaussian fluctuations 
with a variance smaller than the value assumed here. But at their core
the problems would not go away. Observations can only be used to support 
internal or external forcing mechanisms when the stochastic climate 
noise is under control. Otherwise one has to rely on theoretical 
calculations alone.


\end{document}